\documentclass[11pt]{article}
\usepackage{fleqn,cospar}

\newcommand\mydraft[2]{#2}  

\mydraft{}{ \topmargin -20mm }

\mydraft{}{}  

\usepackage{natbib2e}

\input epsf

\usepackage{url}

\newcommand\zzz[2]{#2}  

\renewcommand\citep[1]{(\citealt{#1})}
\newcommand\citepf[1]{(\citealt*{#1})} 

\def\gtapprox{\,\lower.6ex\hbox{$\buildrel >\over \sim$} \, }
\def\ltapprox{\,\lower.6ex\hbox{$\buildrel <\over \sim$} \, }

\def\hMpc{\mbox{h$^{-1}$ Mpc}}

\def\deg{\ifmmode^\circ\else$^\circ$\fi}    

\def\ltapprox{\,\lower.6ex\hbox{$\buildrel <\over \sim$} \, }
\def\ttimes{{\scriptstyle \times}}

\def\centreline{\centerline}

\def\hGpc{\mbox{$h^{-1}$ Gpc}}

\def\Omm{\Omega_{\mbox{\rm \small m}}}
\def\rinj{r_{\mbox{\rm \small inj}}}

\newcommand\mycaptionfont{ }

\def\fptstor{ 
\begin{figure} 
\centreline{\epsfxsize=8cm
\zzz{\epsfbox[42 31 534 527]{"`gunzip -c pts_1.ps.gz"} }
{\epsfbox[42 31 534 527]{"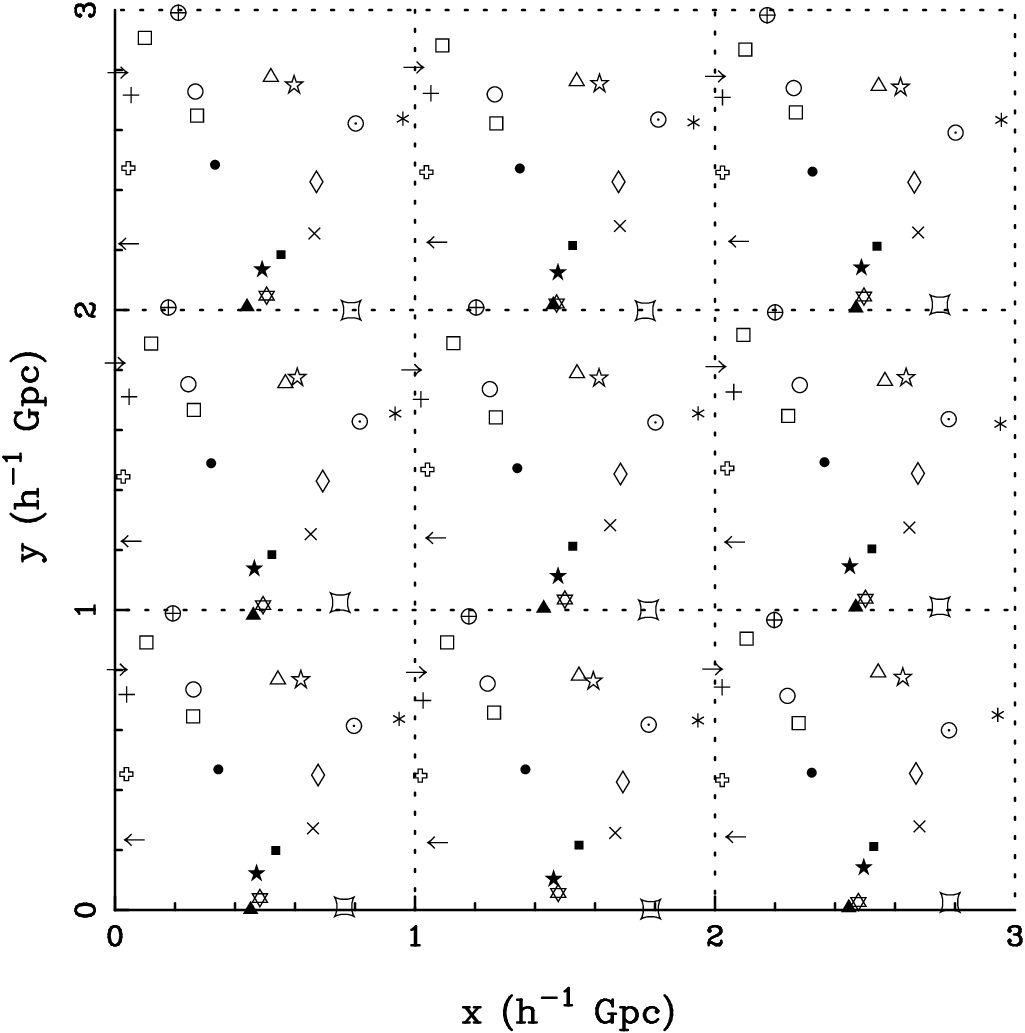"} }
}
\caption{\label{f-ptstor} \mycaptionfont 
Distribution of images of $N=20$ objects at cosmological distances in a multiply
connected two-dimensional $T^2$ model of side length 
$L=1${\hGpc}, in the apparent space (covering space). 
The $N$ objects are placed randomly according to a uniform
distribution in the fundamental domain.
Each physically distinct 
object is shown by a separate symbol. Multiple topological 
{\it images} of a single
{\it object} are each shown by the same symbol. 
Random uniform offsets of $\pm 25${\hMpc} are added in order
that correlations in pair separation plots (below) are easily visible to the
eye. More realistic offsets due to peculiar velocities and/or
measurement error would be of the order of $\sim 1-10${\hMpc} over
these time intervals, i.e. slightly smaller. Dotted lines are used
to guide the eye, but have no physical meaning. This model 
is locally homogeneous and isotropic.
}
\end{figure} 
}

\def\fptskle{ 
\begin{figure} 
\centreline{\epsfxsize=8cm
\zzz{\epsfbox[42 31 534 527]{"`gunzip -c pts_2.ps.gz"} }
{\epsfbox[42 31 534 527]{"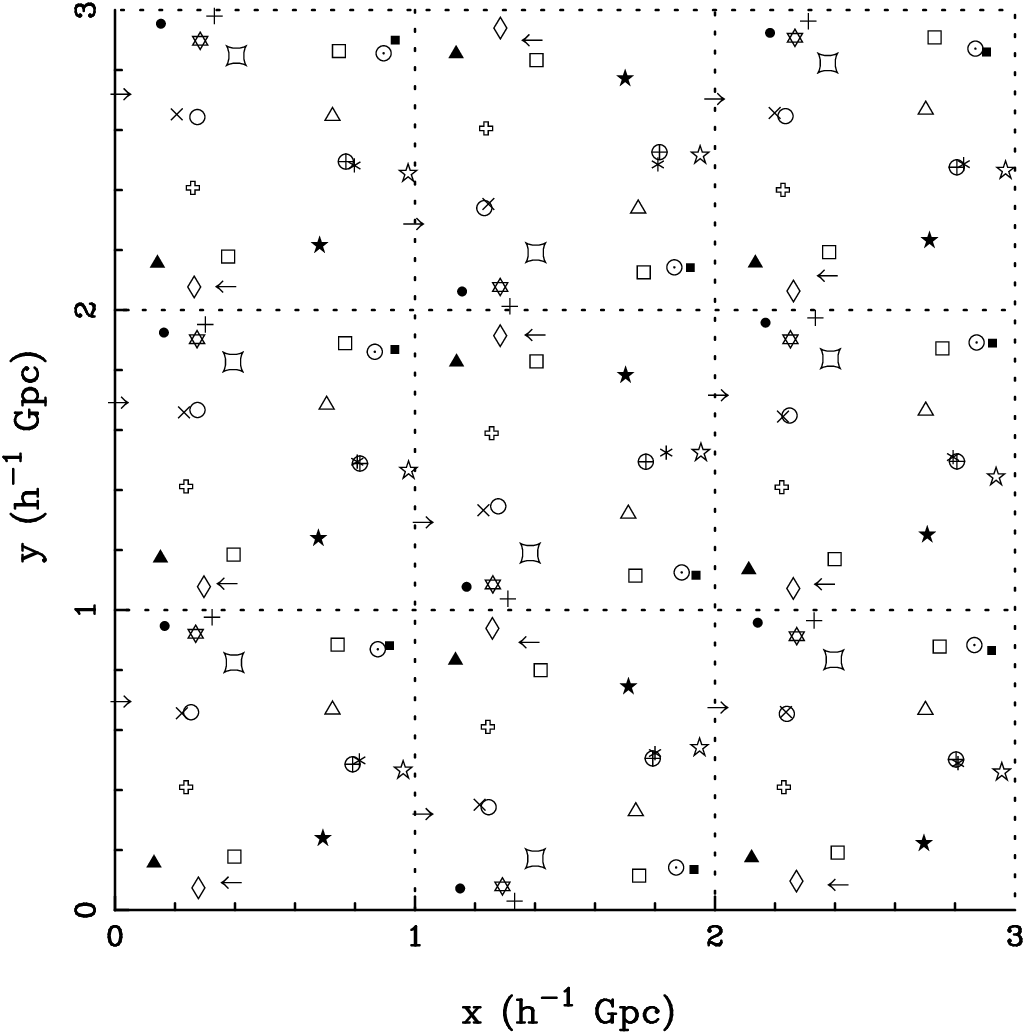"} }
}
\caption{\label{f-ptskle} \mycaptionfont 
As for Fig.~\protect\ref{f-ptstor}, for a Klein bottle model, i.e.
identical to $T^2$ except that before identifying the left and 
right hand sides of the fundamental domain, a twist or reversal
is applied. This model 
is locally homogeneous and isotropic.
}
\end{figure} 
}

\def\fseptor{ 
\begin{figure} 
\centreline{\epsfxsize=8cm
\zzz{\epsfbox[42 31 534 527]{"`gunzip -c sep_1ih1.ps.gz"} }
{\epsfbox[42 31 534 527]{"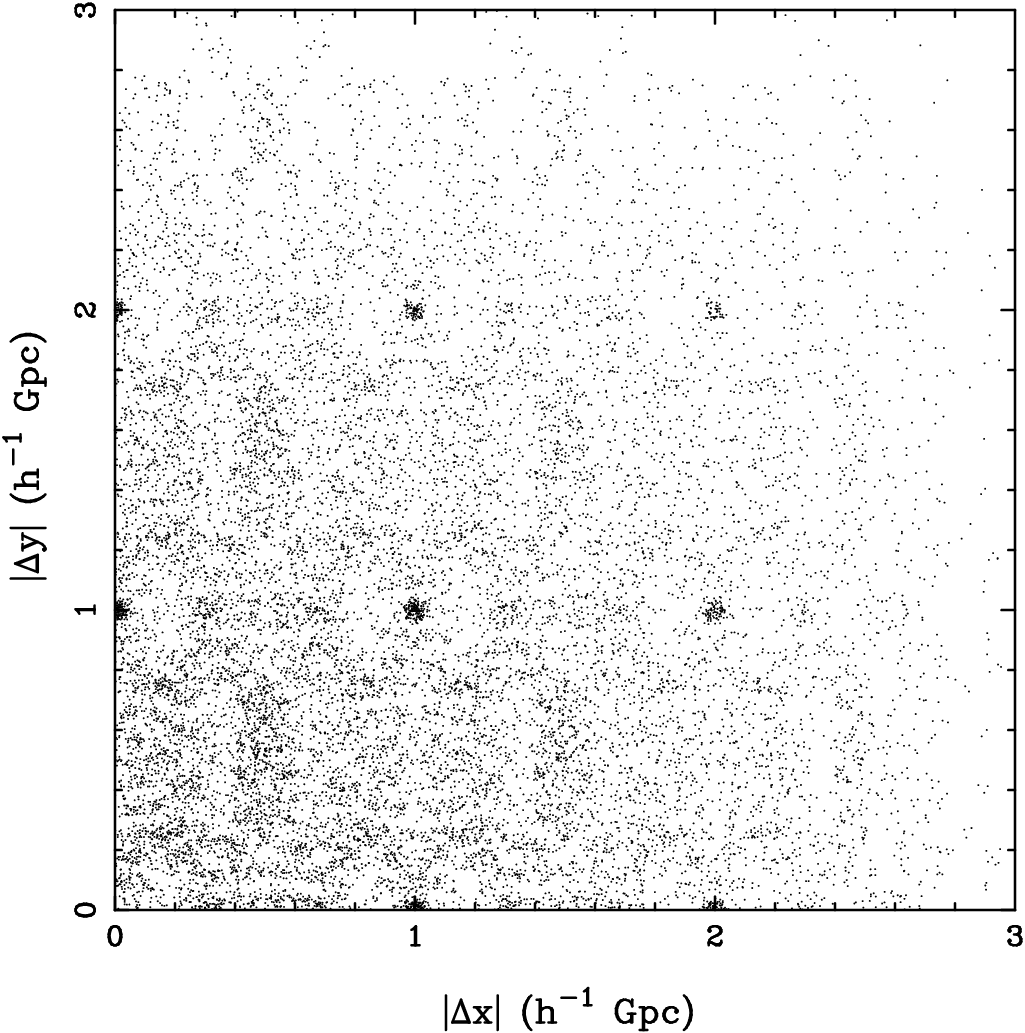"} }
}
\caption{\label{f-septor} \mycaptionfont 
Separation vectors 
$(|x_i-x_j|,|y_i-y_j|)$ for all pairs of images in the $T^2$ model
shown in Fig.~\protect\ref{f-ptstor}. The model is globally 
{\it anisotropic}, since the pattern 
of ``spikes'' at the points 
$\{( i\mbox{\hMpc}, j\mbox{\hMpc}), i,j \in \mbox{\cal N}\}$ 
changes under rotation 
of the $x-y$ axes. The model is globally {\it homogeneous}: the 
pattern is identical for different choices of the origin.
}
\end{figure} 
}

\def\fsepkle{ 
\begin{figure} 
\centreline{\epsfxsize=8cm
\zzz{\epsfbox[42 31 534 527]{"`gunzip -c sep_2ih1.ps.gz"} }
{\epsfbox[42 31 534 527]{"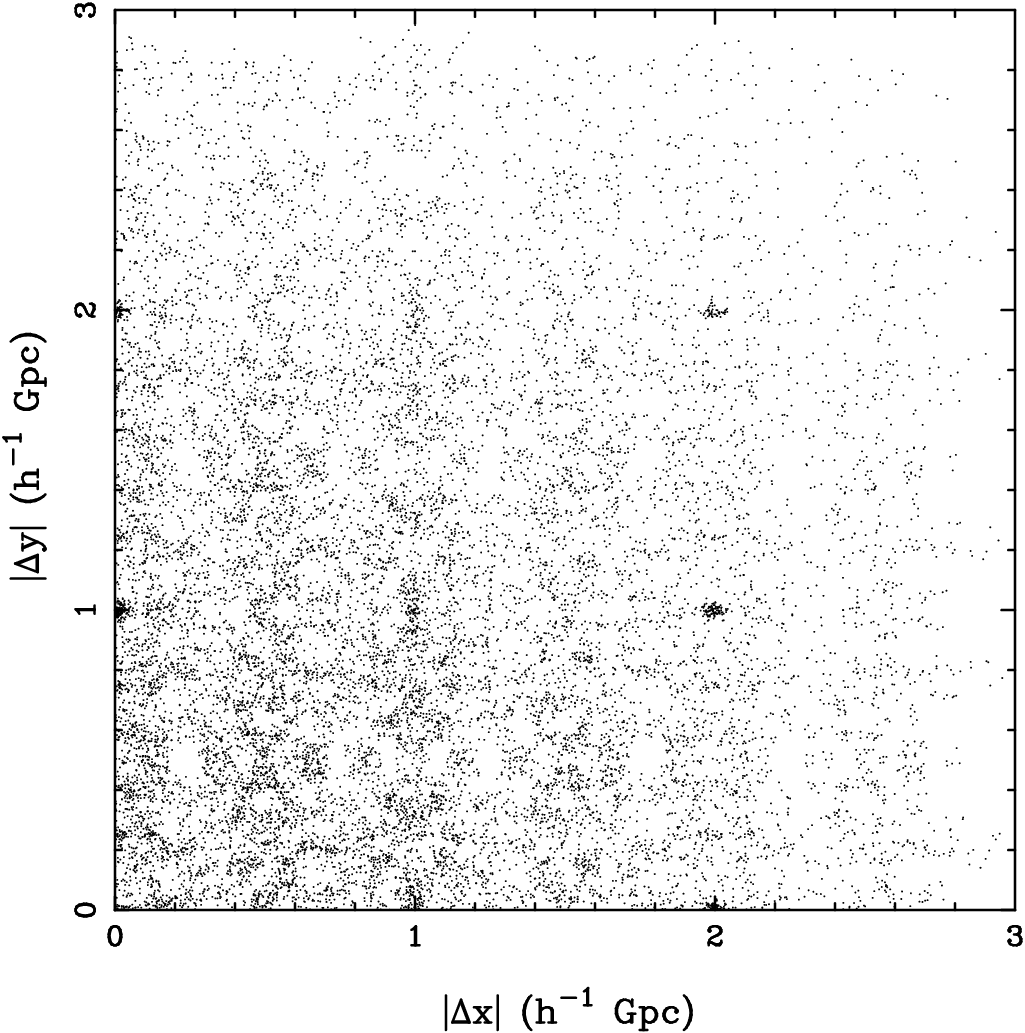"} }
}
\caption{\label{f-sepkle} \mycaptionfont 
Separation vectors 
$(|x_i-x_j|,|y_i-y_j|)$ for all pairs of images in the Klein bottle model
shown in Fig.~\protect\ref{f-ptskle}. 
As for Fig.~\protect\ref{f-ptstor},  
the model is globally 
{\it anisotropic}. Because of the orientation reversal, some of the
separations corresponding to ``cosmic crystallography spikes'' 
\protect\citep{LLL96} present in Fig.~\protect\ref{f-ptstor} 
seem to have disappeared. This is a clue to establishing inhomogeneity.
}
\end{figure} 
}

\mydraft{
\def\fsepkleab{ 
\begin{figure} 
\fsepkleacore  
\end{figure}

\begin{figure} 
\fsepklebcore  
\end{figure}
}}{
\def\fsepkleab{ 
\begin{figure} 
\fsepkleacore  
\fsepklebcore  
\end{figure}
}}

\def\fsepkleacore{
\centreline{\epsfxsize=8cm
\zzz{\epsfbox[42 31 534 527]{"`gunzip -c sep_2ih2.ps.gz"} }
{\epsfbox[42 31 534 527]{"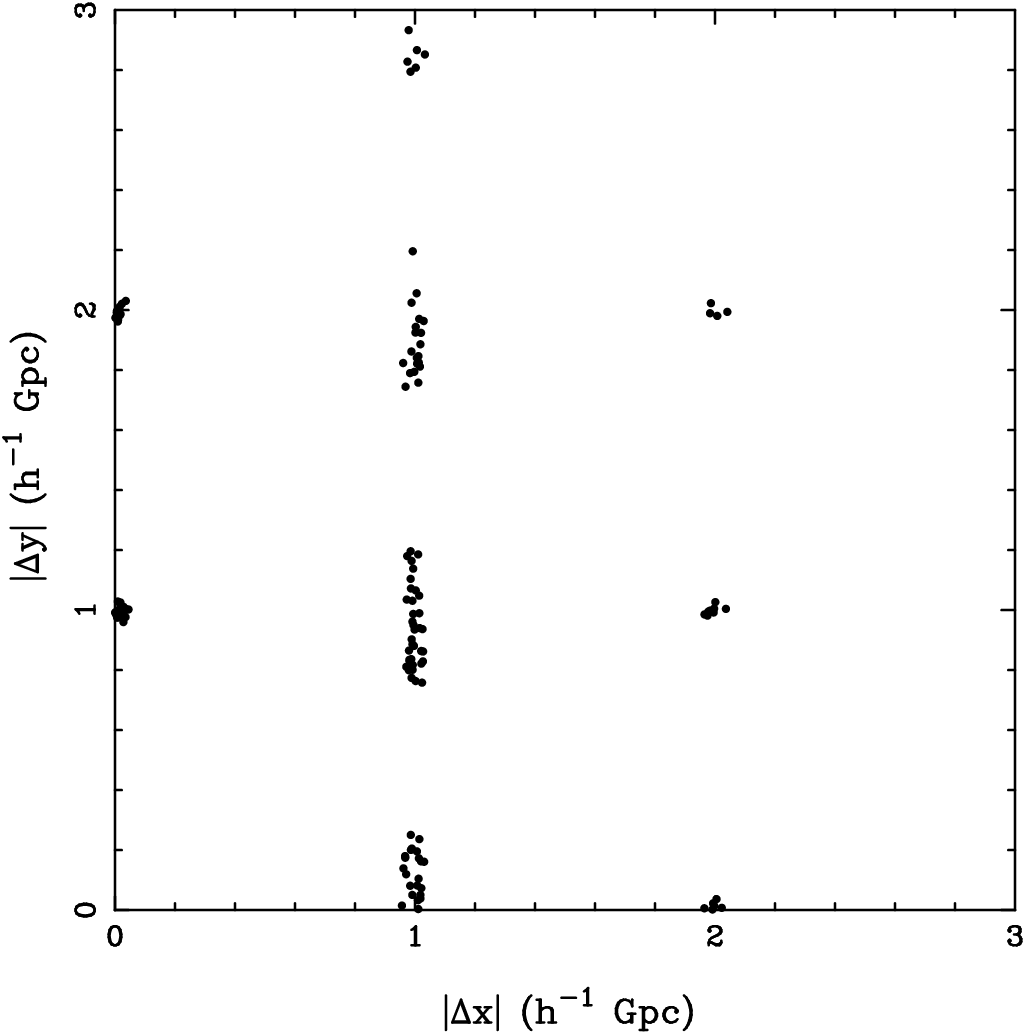"} }
}
\caption{\label{f-sepklea} \mycaptionfont 
Klein bottle image pair separations, as for Fig.~\protect\ref{f-ptskle},  
but only showing pair separations for pairs for which the
two multiple images are those of a single physical object, 
where the $y$ coordinate of an element of the pair satisfies 
$ | (y-y_0) +nL/2| < L/8 $, for some $n\in Z$, and where
$L=1${\hGpc} and $y_0=0${\hGpc}.
One element of a pair of images of an object satisfies this relation
if and only if the other satisfies it. In words, these are the objects
which are close to either the horizontal 
``edges'' or to the half-way dividing lines between the horizontal 
``edges''. 
Some of the ``cosmic crystallography spikes'' 
present in Fig.~\protect\ref{f-septor} but missing from 
Fig.~\protect\ref{f-sepkle} have now reappeared, but are 
spread out in the vertical direction.
}
} 

\def\fsepklebcore{
\centreline{\epsfxsize=8cm
\zzz{\epsfbox[42 31 534 527]{"`gunzip -c sep_2ih3.ps.gz"} }
{\epsfbox[42 31 534 527]{"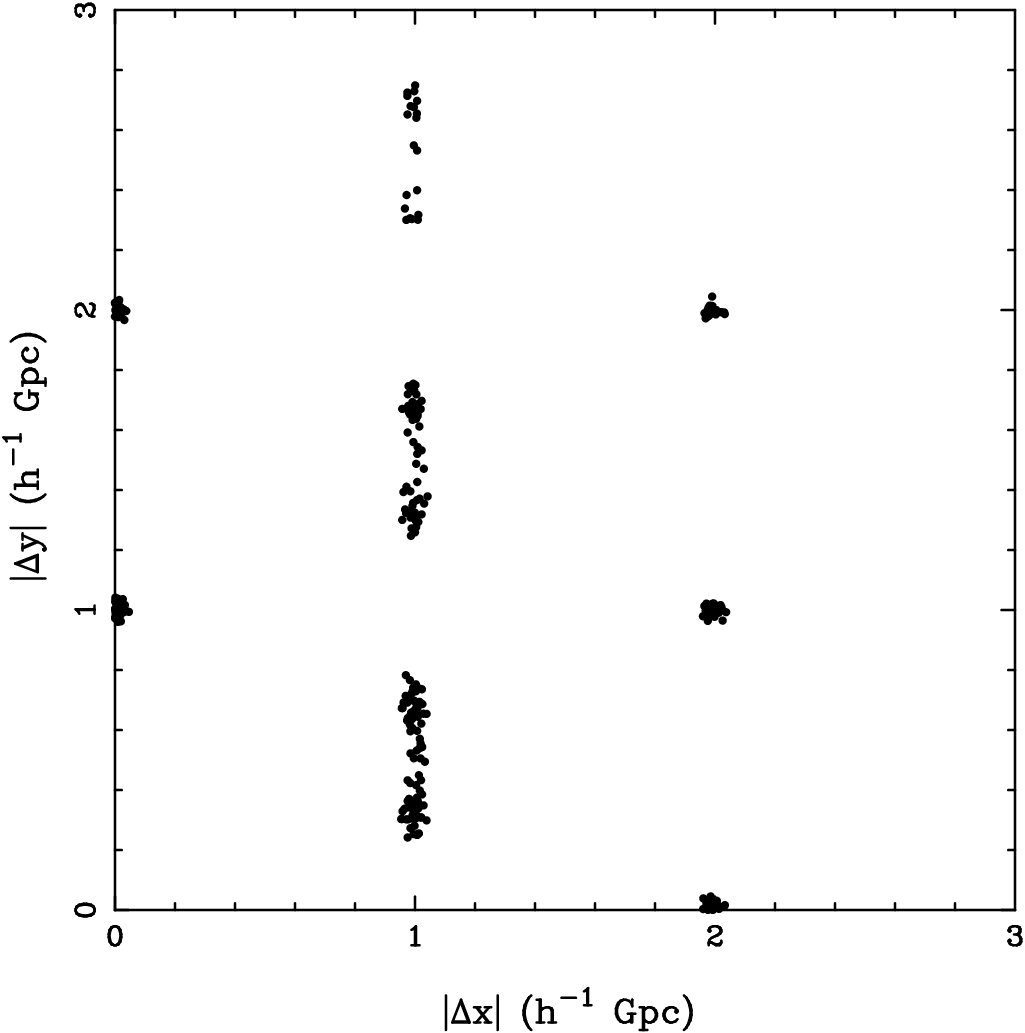"} }
}
\caption{\label{f-sepkleb} \mycaptionfont 
Klein bottle image pair separations, as for Fig.~\protect\ref{f-sepklea},  
where the $y$ coordinate of an element of the pair satisfies
$ | (y - y_0) +nL/2| < L/8 $, for some $n\in Z$, 
and where $L=1${\hGpc}, but $y_0=L/4=0.25${\hGpc}.
These pairs are the complement to 
the set in Fig.~\protect\ref{f-sepklea}. The difference between
Figs~\protect\ref{f-sepklea} and \protect\ref{f-sepkleb}
shows that the Klein bottle is inhomogeneous: 
for different choices of the zero-point $y_0$, the figure changes.
}
}  

\title{
{\mbox{\normalsize \em Advances in Space Research (in press) 
\rule{13cm}{0cm}\rule{0cm}{3cm}}}
\\{\ }
\\HOW TO AVOID THE AMBIGUITY IN APPLYING THE COPERNICAN PRINCIPLE
FOR COSMIC TOPOLOGY: TAKE THE OBSERVATIONAL APPROACH}

\author{B. F. Roukema\address{Inter-University Centre 
for Astronomy \& Astrophysics,
Post Bag 4, Ganeshkhind, Pune 411007, India}
\address{DARC, Observatoire de Paris-Meudon, 5, place Jules Janssen,
Meudon Cedex, France\\
{ email: boud.roukema@obspm.fr}} 
}

\begin{document}

\maketitle

\begin{abstract}
It is often stated that homogeneity and isotropy of the Universe are 
assumptions of the almost Friedmann-Lema\^{\i}tre (FL) model (the hot
big bang model), inspired from the Copernican Principle. However, only 
{\it local} homogeneity and isotropy are required by the model: 
multiply connected almost FL models are locally homogeneous and
isotropic, but they can be {\it globally} anisotropic and/or 
{\it globally} inhomogeneous. Toy models are used here to show
how global anisotropy and/or global inhomogeneity of an almost FL model 
could be shown directly in observations. This approach may avoid 
having to make any assumptions regarding global anisotropy and
inhomogeneity.  
\end{abstract}

\section*{INTRODUCTION}

The Copernican principle \citep{Coper43}, defined for the present paper 
as the principle according to 
which the observer should not happen to be located in any special location
in the Universe, is generally {\it applied} in one of three ways in twentieth
century cosmology:
\begin{list}{(\roman{enumi})}{\usecounter{enumi}}
\item by assuming the Cosmological Principle, in which spatial sections 
of the Universe at constant cosmological time are close to homogeneous
and isotropic when averaged on ``large enough'' length scales;
\item by assuming the Perfect Cosmological Principle, in which all of 
space-time is close to homogeneous and isotropic, on average; or
\item by assuming that the Universe is very chaotic, very far from 
homogeneity, but that by the Weak Anthropic Principle, we happen to live
in a tiny bubble which is inflated to a locally close to flat 
Friedmann-Lema\^{\i}tre model;
\end{list}
or it is {\it rejected} 
\begin{list}{(\roman{enumi})}{\usecounter{enumi}} 
\addtocounter{enumi}{3}
\item by taking the ``observational'' point of view that the observer
lives at the centre of a close to isotropic but radially very 
inhomogeneous three-dimensional 
space-time cone with a spherical ($S^2$) boundary defined
by the age of the local patch of the Universe, so that no space outside
of this sphere exists as far as science is concerned, and which 
happens to coincide with what one expects in case (i) above.
\end{list}

Case (i) underlies what is generally accepted as the standard hot
big bang model, which can be more formally referred to 
as the almost Friedmann-Lema\^{\i}tre model 
(e.g. \citealt{Schw00,Schw98,deSitt17,Fried23,Fried97,Lemait58,Wein72}).
This model may or may not have zero curvature, and it may or may not
have trivial topology. Numerous observations 
now knit together to provide a very solid foundation
for the almost Friedmann-Lema\^{\i}tre model, 
within the spatial, temporal and density limits of its validity.

Case (ii), that of a steady state or of a quasi-steady-state model
\citepf{HBN93}, is only defined in its general outline, and would require
many more close links with recent observations in order to have 
predictive power. For example, the case (i) interpretation of 
quasars' redshifts as primarily due to expansion of the Universe,
when applied to observations,  
implies that quasars and Lyman break galaxies 
at high redshift trace large scale structure at 
$L\sim130\pm10${\hMpc} (comoving) and that they 
function as a standard ruler which implies
a low matter density $\Omm \sim 0.3$ \citep{RM00a,RM00b,BJ99}. 
This would be very difficult to explain if quasar (and Lyman break
galaxy) redshifts were non-cosmological.

Case (iii) corresponds to the presently very popular chaotic inflation
scenarios. Whether or not the Weak Anthropic Principle is consistent with
the Copernican Principle is out of the scope of this paper.

Case (iii) is generally considered an extrapolation (towards short 
cosmological times and large spatial distances) of the almost 
Friedmann-Lema\^{\i}tre (FL) model. Case (ii) could also conceivably be 
defined as an extension of the FL model.

The anti-Copernican case (iv) is essentially identical to case (i) 
as far as observational analysis is concerned --- as long as the 
topology of the Universe remains unmeasured. However, theoretical 
calculation is somewhat difficult if case (i) is not resorted to.


If the topology of space were measured by observation 
(\citealt{LR99}, and see below), then the size of the Universe would be
finite and less than the horizon diameter in at least one direction,
i.e. finite.
Models of case (ii) would then require serious revision and would no longer
satisfy the Perfect Cosmological Principle, and case (iii) models 
would presumably 
require individual ``universes'' to be born as separate space-times
and might have difficulty surviving Occam's Razor.

Case (iv) would also be partly or totally avoidable: only the 
{\it apparent} space would be a sphere centred on the observer.
At least some parts of the ``edge of the Universe'' of the case 
(iv) interpretation would no longer be edges.

On the other hand, whether or not case (i) would remain valid 
if the Universe is found to be multiply connected 
would depend on the resolution of the ambiguity in the meanings 
of the words ``homogeneous'' and ``isotropic''. The solution of the 
Einstein-Hilbert equations is only found as a local solution, so only
{\it local} homogeneity and isotropy are required. In general, 
although Friedmann-Lema\^{\i}tre models satisfy 
{\it local} homogeneity and isotropy, they need not satisfy 
{\it global} homogeneity and isotropy. Observational support 
for local homogeneity and isotropy does not imply 
global homogeneity and isotropy. 

In this paper, the relevance of different interpretations of these
two words in the context of the global geometry 
of space (curvature and topology) 
is explained and illustrated by simple toy models in two dimensions. 
By attempting to measure 
the more subtle elements of inhomogeneity 
and anisotropy which would reveal the global shape 
of the Universe, some of the ambiguity in deciding how to apply 
the Copernican principle to observational cosmology 
could be avoidable: empirical results sometimes help resolve
theoretical dilemmas.

Proper distance (\citealt{Wein72}, Eq.~14.2.21) in comoving units 
is used throughout this paper unless otherwise stated. 
The Hubble constant is parametrised here 
as $h\equiv H_0/100\,$km~s$^{-1}$~Mpc$^{-1}.$ 
Values of the density parameter, $\Omm$, 
the dimensionless cosmological constant,
$\Omega_\lambda$, the dimensionless curvature, 
$\Omega_\kappa\equiv \Omm + \Omega_\Lambda -1$, and the
curvature radius 
$R_C \equiv (c/H_0) \Omega_\kappa^{-1/2},$
are indicated where used.

\section*{COSMIC TOPOLOGY, HOMOGENEITY AND ANISOTROPY}

\subsection*{Local Homogeneity and Isotropy}

When justifying the
assumptions of (i) homogeneity and (ii) isotropy, these are generally interpreted
to mean that 
\begin{list}{(\alph{enumi})}{\usecounter{enumi}}
\item the comoving spatial density of astrophysical objects 
is approximately constant, and 
\item the solid angular distribution of 
distant objects (or microwave background fluctuations) is 
approximately constant, after correction for the very anisotropic biases
due to living in a dusty disk galaxy and due to our peculiar velocity 
with respect to the comoving reference frame,
\end{list}
respectively.

Both are reasonably consistent with numerous observations, and the 
latter has been further reinforced recently 
by the only cosmological class of objects known to be isotropic {\it in spite}
of galactic absorption and our peculiar velocity: the gamma ray
bursts. These may in fact reveal the births of black holes, i.e. of 
regions of extreme curvature (inhomogeneity) on very small scales, that are
briefly visible before being engulfed by 
event horizons \citep{JDM00}.

These two properties (a), (b) are generally referred to 
as ``local homogeneity and isotropy''. Solutions to the Einstein-Hilbert 
equations are normally only found as local solutions of differential
equations, and so only local assumptions of homogeneity and isotropy 
have any effect on Friedmann-Lema\^{\i}tre models.

One might also refer to these as first order homogeneity and 
isotropy, since only one-point statistics of density are considered.

\subsection*{Cosmic Topology and Global Anisotropy}

For the earliest known article on cosmic topology, see 
\citet{Schw00,Schw98}. A recent review 
is that of \citet{LR99}. The multiply connected 
models which have been compared with observations are all FL models,
i.e. they are {\it locally homogeneous and isotropic}.

A simple way to illustrate multiply connected 3-manifolds which may 
correspond to the comoving 3-space we live in is to start with the
3-torus, $T^3$. This can be thought of 
\begin{list}{(\roman{enumi})}{\usecounter{enumi}}
\item {\it physically}, as a cube of which opposite faces are identified 
so that space is continuous, with no boundaries, but finite,
\item {\it observationally}, as a tiling of Euclidean space $R^3$ 
by multiple copies of the single {\it physical} ``cube'', or
\item as the analogy one dimension higher of 
a 2-torus embedded in $R^3$ but endowed with an intrinsically flat
metric, i.e. a metric different to the normal metric of $R^3$.
\end{list}

For a demonstration of basic properties of locally homogeneous and
isotropic manifolds, it is useful to subtract one dimension, i.e. 
to consider 2-manifolds. This is the approach adopted here
for pedagogical purposes. For the physical Universe, the results 
extrapolate to the three-dimensional case.

\fptstor
\fseptor

The cube (i) is referred to as a {\it fundamental domain\/}, and 
the space (ii) is referred to as the {\it universal 
covering space} or in a new coinage, {\it apparent space}. 
Both are equivalent ways of thinking about the same manifold. 
The former is useful for conventional physical reasoning, according
to which any particle or object only exists at one point in a space, but
difficult to use for interpreting observations.
The latter is useful for observational analyses, but surprising and sometimes
confusing for physical reasoning, since objects ``exist'' many times in it.

In two dimensions, the corresponding space is the flat 2-torus 
$T^2$, of which the fundamental domain is a square (in general a 
parallelogram, but a square is used for the present paper), and the
universal covering space is $R^2$.

Figure~\ref{f-ptstor} shows what such a universe would look like. 
For reference, the reader should remember that the horizon size 
for likely values of the curvature parameters 
$(\Omm\approx 0.3, \Omega_\Lambda\approx 0.7)$ is $R_H \approx 10${\hGpc},
and that objects like quasars or ``Lyman break galaxies'' are 
seen up to typical redshifts of $z\sim2-4$, which correspond to 
roughly half this distance. Clusters of galaxies and galaxies are
mostly only seen to a few $100${\hMpc}.

Present constraints on the size of the Universe (defined here as the 
injectivity diameter $2\rinj$, which is the length of the
shortest spatial geodesic linking an object to an image of itself) 
are $2\rinj \gtapprox 1${\hGpc}, whether the Universe is hyperbolic,
flat or spherical. So, in order to see multiple images 
of a single object, catalogues of objects which 
are seen to typically a few tenths of $R_H$ would be required. This 
implies that 
effects of aging of the objects, or of viewing angles, or of movement
relative to the comoving reference frame, would make 
multiple images of a single object appear quite different, so that proving
the identity of two images would be difficult.

So, the identity between multiple copies of symbols
in Fig.~\ref{f-ptstor} would not be obvious in any category
of presently known objects. 

The model is {\it locally homogeneous and isotropic}: 
any study in a small region around a point is independent of the
orientation of the axes of a chosen coordinate system and of the origin of
that system.


{\it The model of Fig.~\ref{f-ptstor} is globally homogeneous, but globally 
anisotropic.}

\fptskle
\fsepkle

\fsepkleab

The anisotropy is shown by Fig.~\ref{f-septor}, which shows vectors
separating all pairs of images in the simulated universe. Separations
between multiple images of single objects correspond to multiples
of the generators [the vectors $(0,1)$ and $(1,0)$] and clearly cause
clustering around these points. As pointed out by \citet*{LLL96}, 
a one-dimensional pair separation histogram 
$\Delta N ( \Delta r )$ should show sharp spikes corresponding to 
these cluster points. (If no offsets were added between multiple images,
then the spikes would be Dirac $\delta$ functions.)

The pairs of images contributing to these spikes can be referred
to as {\it generator pairs} or as {\it Type II pairs} \citepf{ULL99}, 
which induce spikes in flat multiply connected spaces, but not
in curved multiply connected spaces \citep{Gomero99a,ULL99}. 

If the $x-y$ axes are chosen at a different angle, then the diagram
would be rotated: the pattern of spikes is not invariant (i.e. varies) 
under rotation. This proves global anisotropy. 

However, the diagram would be statistically identical under a shift
(without rotation) of the origin: the space is homogeneous. 

\subsection*{Cosmic Topology and Global Inhomogeneity}

\subsubsection*{Use of ``generator pairs''}

Many 3-manifolds are globally inhomogeneous, but the effect is observationally
quite subtle.

The effect is illustrated here for the Klein bottle. This is like $T^2$,
except that a twist is applied 
before identification of one pair of edges 
(here, the ``vertical'' edges). The distribution of images is shown 
in Fig.~\ref{f-ptskle}. Again, this model is 
{\it locally homogeneous and isotropic}. 

Fig.~\ref{f-sepkle} shows that, as for the $T^2$ model, the Klein bottle
is {\it globally anisotropic}: 
rotation of the coordinate axes would modify the pattern
of crytallographic spikes.

In order to show global inhomogeneity using ``generator pairs'', 
\begin{list}{(\roman{enumi})}{\usecounter{enumi}}
\item knowledge of which pairs of images
are due to multiple images of single objects is required, and
\item this knowledge is required {\it despite the fact that the 
crystallographic spikes corresponding to ``inhomogeneous generator pairs'' 
are smeared out making them very difficult to detect.}
\end{list}

Both (i) and (ii) are, of course, easy for a simulated
universe: in the observed Universe they are difficult.

Comparison of Figs~\ref{f-septor} and \ref{f-sepkle} shows that some
crystallographic spikes seem to disappear due to the twist in identifying
the vertical edges of the fundamental domain.

Figures~\ref{f-sepklea} and \ref{f-sepkleb} show where these spikes
have ``disappeared'' to, using property (i). Different choices of 
the $y$ origin (see figure captions) yield different patterns.
Together, the two figures show how the spikes have been smeared out.

Hence, the Klein bottle is {\it globally inhomogeneous}.

One way to apply this technique observationally would be 
\begin{list}{(\alph{enumi})}{\usecounter{enumi}}
\item to use properties of
galaxies or quasars which uniquely label and identify individual
objects, despite the large fractions of the age of the Universe over
which the objects age. 
\end{list}
This seems difficult given present observational
limitations and known techniques, but not impossible.

An alternative method would be 
\begin{list}{(\alph{enumi})}{\usecounter{enumi}} \addtocounter{enumi}{1}
\item firstly to detect the crystallographic 
spikes of generator pairs 
which {\it do} appear, and which would imply a fundamental domain 
of size $2L=2${\hGpc} in the horizontal direction, 
and secondly to hypothesise that the real fundamental
domain is just half this length. 
\end{list}
This would make predictions of which pairs
of objects should correspond to those seen separately in 
Figs~\ref{f-sepklea} and \ref{f-sepkleb}. If the predicted multiply 
imaged pairs
really were identical, then proving this by observation 
would be much easier than the ``needle in a haystack'' 
requirement of finding the corresponding pairs 
blindly as in the previous suggestion (a).


\subsubsection*{The Copernican principle in an inhomogeneous universe}

Figs~\ref{f-sepklea} and \ref{f-sepkleb} illustrate one way in which
the Copernican principle could be applied directly. One would not expect
the observer to be at any special value of $y_0$: the value 
$y_0=0.00\pm0.01${\hGpc} for the toy model shown here would be surprising.

\subsubsection*{Use of ``local pairs'' or ``local $n$-tuplets''}

Inspection of Fig.~\ref{f-ptstor} and a little reflection show
that generator pairs are not the only pairs of images which occur
repeatedly in a separation vector plot (or separation distance
histogram).

Close pairs of images of {\it non}-identical objects within a single
fundamental domain also define 
separation vectors which are repeated in the apparent space, though
in a different way to generator pairs. These are {\it local pairs} 
or {\it Type I} pairs (\citealt{Rouk96}; Section 2 of \citealt{ULL99}).
Since there are nine copies of the
fundamental domain shown in Fig.~\ref{f-ptstor}, there are 
nine local pairs clustered around each of the 
$20\ttimes 19/2=190$ separation vectors of modulus smaller than 
$L=1${\hGpc}. This is visible as a ``graininess'' in 
Figs~\ref{f-septor} and \ref{f-sepkle}.

In a flat space, the same effect also occurs for 
pairs across many copies of the fundamental domain, but in general does
not occur in curved spaces. Thus the term ``local'': 
it is most useful just to consider relatively close pairs of images.

\citet{FagG99a} noticed the effect of the local pairs in pair
separation histograms, and \citet{ULL99} showed how to collect these
together and detect multiple connectedness. If these led to detection
of multiple connectedness for a model known (mathematically)
to be inhomogeneous, then this would enable (a) above to be applied:
the generator pairs could be used to observationally illustrate
inhomogeneity. However, local pairs in themselves would not 
show inhomogeneity: they are local.

Similarly, local $n$-tuplets \citep{Rouk96} would not directly show
inhomogeneity. However, since non-orientable manifolds are inhomogeneous,
the discovery of significant numbers of matching $n$-tuplets, where 
some of these required orientation reversals, would imply
inhomogeneity.

For more details of global homogeneity and isotropy questions of 2-manifolds,
and more particularly of 3-manifolds, see \citet{Thur97}.

\section*{CONCLUSION}

The Copernican Principle, when used to motivate {\it local} homogeneity and
isotropy of the Universe, does not necessarily imply {\it global} homogeneity 
and isotropy. If the topology of the Universe is observable, then 
this may reveal 
\begin{list}{(\roman{enumi})}{\usecounter{enumi}}
\item that the Universe is globally anisotropic, though 
locally isotropic and both globally and locally homogeneous, as in the
case of a $T^2$ model in two dimensions, or 
\item that the Universe is globally both anisotropic and 
inhomogeneous, even though 
locally isotropic and homogeneous, as in the
case of a Klein bottle model in two dimensions.
\end{list}

Most techniques for measuring cosmic topology should easily lead 
to direct geometrical illustrations of 
global anisotropy if multiple connectedness is detected 
(unless the 3-manifold we live is globally isotropic: 
the projective space, $S^3/Z_2$, which can be thought of 
by identifying all opposite points of the hypersphere $S^3$, is
globally both isotropic and homogeneous, even though multiply
connected). 

By contrast, global 
inhomogeneity in a flat multiply connected manifold would be
difficult to detect directly. However, prior knowledge of 
multiple connectedness would help in obtaining
direct illustrations of global inhomogeneity.

Present observations imply that the observable Universe is close 
to flat \citep{Boom00a,Maxima00a,RM00a} and that multiply connected
flat models of size $2\rinj \sim 2R_H/10$ are consistent with 
present large angle microwave background data \citep{Rouk00c}.
The observational approach to global geometry
might just possibly provide 
an empirical solution to an otherwise philosophical question, by 
showing whether not both local {\it and} global homogeneity and/or
isotropy are valid assumptions..


\newcommand\joref[6]{#1, #6, \newblock {\it #2 }{\bf #3, } #4, #5.}  
\newcommand\confref[5]{#1, #5, {#2 }{#3, } #4.} 
\newcommand\inpress[5]{#1, #5, #2 in press}  
\newcommand\prepr[4]{#1, #3.  #2 (preprint)} 
\newcommand\epref[4]{#1, #4, \newblock {\it #2}, #3.}

\def\apj{Astrophys. J.} 
\def\apjs{ApJS}                 
\def\aj{AJ}                       
\def\aap{Astron. \& Astroph.} 
\def\aaps{A\&AS}            

\def\mnras{Mon. Not. R. Astr. S.}
\def\araa{ARA\&A}
\def\cqg{Class. \& Quant. Grav.}   

\end{document}